\documentclass[times, 10pt,onecolumn]{article}
\usepackage{amsmath, amssymb, enumerate}

\usepackage{fancybox}
\usepackage{amsfonts}
\usepackage[usenames]{color}
\usepackage{listings}
\usepackage{hyperref}
\usepackage[linesnumbered,vlined]{algorithm2e}
\usepackage{graphicx}
\usepackage{times}
\usepackage{psfrag}
\usepackage{subfigure}
\usepackage{caption}
\usepackage{multirow}
\usepackage{epsfig}
\usepackage{url}

\usepackage{color}
\def\fixme#1{\typeout{FIXED in page \thepage : {#1}}
\bgroup \color{red}{} \egroup}

\usepackage{rotating,tabularx}

\makeatletter
\def\url@leostyle{%
\@ifundefined{selectfont}{\def\UrlFont{\sf}}{\def\UrlFont{\small\ttfamily}}}
\makeatother

\newcommand{\bottomrule}{\hline}
\newcommand{\toprule}{\hline}
\newcommand{\midrule}{\hline}
\title{Protecting Memory-Performance Critical Sections in Soft Real-Time Applications}
\author{Heechul Yun$^\dagger$, Santosh Gondi$^\ddagger$, Siddhartha Biswas$^\dagger$\\
$^\dagger$ University of Kansas, USA. \{heechul.yun,sid.biswas\}@ku.edu\\
$^\ddagger$ Bose Corporation, Framingham, MA, USA. santosh\_gondi@bose.com
}
\date{}

\begin{document}
\maketitle
\thispagestyle{empty}
\begin{abstract}
Soft real-time applications such as multimedia applications often
show bursty memory access patterns---regularly requiring a high
memory bandwidth for a short duration of time. Such a period is often
critical for timely data processing. Hence, we call it a
memory-performance critical section. Unfortunately, in multicore
architecture, non-real-time applications on different cores may also
demand high memory bandwidth at the same time, which can substantially
increase the time spent on the memory performance critical sections.

In this paper, we present BWLOCK, user-level APIs and a memory
bandwidth control mechanism that can protect such memory performance
critical sections of soft real-time applications. BWLOCK provides
simple lock like APIs to declare memory-performance critical
sections. If an application enters a memory-performance critical
section, the memory bandwidth control system then dynamically limit
other cores' memory access rates to protect memory performance of the
application until the critical section finishes.

From case studies with real-world soft real-time applications, we
found (1) such memory-performance critical sections do exist and are
often easy to identify; and (2) applying BWLOCK for memory critical
sections significantly improve performance of the soft real-time
applications at a small or no cost in throughput of non real-time
applications.
\end{abstract}


\section{Introduction}


In a multicore system, an application's performance running on a core
can be significantly affected by other applications on different cores
due to contention in shared hardware resources such as shared Last-Level
Cache (LLC) and DRAM. When the shared resources become bottlenecks,
traditional CPU scheduling based techniques such as raising
priorities~\cite{lehoczky1989rate} or using CPU reservation based approaches
~\cite{faggioli2009implementation,kato2010ecrts,abeni1998rtss}
do not necessarily improve performance of the real-time applications.

In hard real-time systems, such as avionics systems, one solution
adopted in the industry has been disabling all but one core in the
system~\cite{kotaba2013multicore} to completely eliminate the shared
resource contention problem, thereby being able to be
certified~\cite{faa2014certification}~\footnote{The current standard
for certification is designed for unicore
systems~\cite{ARINC653}}. Another approach, adopted in PikeOS, is a
time partitioning technique in which only one core is allowed to
execute for a set of pre-defined time windows~\cite{fisher2013pikeos}.

In the context of soft real-time
systems, on the other hand, a certain degree of performance variation due to
interference is often tolerable. Furthermore, modern multicore
architecture provides a significant amount of parallelism in the
processor architecture (e.g., out-of-order core design) and the memory
subsystems (e.g., non-blocking caches and multi-bank DRAM) that can
absorb a considerable degree of concurrent accesses without noticeable
performance impacts~\cite{hennessy2011computer}. Therefore, it is
highly desirable to develop a solution that can provide better
real-time performance while still allowing concurrent
executions to leverage the full potential of multicore.

In our previous work, we developed a software based memory access
control system, called MemGuard, that allows concurrent memory accesses
from multiple cores---up to certain limits---by providing a minimum
memory bandwidth guarantees to each core in the
system~\cite{yun2013rtas}. One problem of this approach is,
however, that the reservable amount of bandwidth is very small,
compared to the peak memory bandwidth. While it tries to maximize
performance via a prediction based bandwidth reclaiming, accurate
prediction is challenging, especially for bursty memory access
patterns, which are commonly found in many soft real-time
applications.

In this paper, we present BWLOCK, a user-level API and memory
bandwidth control mechanism to protect performance of soft
real-time applications such as multimedia applications.
Our key observation is that interference is most visible when multiple
cores have high memory demands at the same time. In such cases, all
participating cores will be delayed due to queueing and other issues
that cannot be hidden by the underlying hardware. Therefore, in order
to protect performance of real-time applications, the system must
avoid overload situations when the real-time applications have high
memory performance by executing memory intensive regions of the
code. We call such a region as \emph{memory-performance critical
section}.
Fortunately, in many soft real-time applications, such as multimedia
applications, such memory-performance critical sections are often easy
to identify via application level profiling techniques. For example, using
\emph{perf} in Linux, one can identify functions that have very high
memory demands.

Motivated from the observations, BWLOCK provides a lock like API with
which programmers can describe certain sections of code that are
memory-performance critical. When a memory-performance critical
section is being executed, BWLOCK limits the amount of allowed memory
traffic from the other cores to avoid overloading memory
performance. In cases that modifying source code or profiling is not
desired or possible, BWLOCK also allows to declare the entire execution
of an application as memory
performance critical so that whenever the application is scheduled on
a CPU core, its memory performance can be ensured.
We call the former as fine-grained bandwidth locking and the latter as
coarse-grained bandwidth locking.

We applied BWLOCK in two real-world soft real-time
applications---Mplayer and WebRTC framework (as part of the
chromium-browser)---to protect their real-time performance in the
presence of memory intensive non real-time applications. In the case of Mplayer,
we achieve near perfect isolation for the Mplayer at a cost of 17\%
throughput reduction of the non real-time applications in the
coarse-grain mode, or achieve 17\% better real-time performance for
the Mplayer at the cost of only 7\% throughput reduction of the
non-real-time applications in the fine-grain mode. Similar
improvements are observed for WebRTC as well.

Our contributions are as follows:
\begin{itemize}
\item We propose an OS mechanism and API that can substantially
improve performance of soft real-time applications in a
multi-programmed environment such as cloud systems.
\item We present extensive evaluation results using real-world soft
real-time applications demonstrating the viability and the
practicality of the proposed approach.
\end{itemize}

The remaining sections are organized as follows:
Section~\ref{sec:motivation} provides background on software based memory
access control technique and motivating experiments.
Section~\ref{sec:bwlock} presents the design and implementation of
BWLOCK. Section~\ref{sec:setup} describes the evaluation
platforms and the implementation overhead analysis.
Section~\ref{sec:casestudy} presents case study results using two
real-world soft real-time applications.
Section \ref{sec:discussion} discusses limitations and possible
improvements.
We discuss related work in Section \ref{sec:related} and conclude in
Section~\ref{sec:conclusion}.

\section{Background and Motivation} \label{sec:motivation}

In ~\cite{yun2013rtas}, we proposed a software based
memory bandwidth management system called MemGuard in Linux
kernel. The key idea is to periodically monitor and
regulate the memory access rate of each core using per-core hardware
performance counters. If, for example, a group
of tasks generates too much memory traffic and delays the
critical real-time tasks, MemGuard can regulate the memory access rates of
the cores running the offending tasks. With the regulation mechanism,
MemGuard offers a bandwidth reservation service that partitions a
fraction of available memory bandwidth among the cores and ensures the
reserved bandwidth is to be guaranteed all the time. The bandwidth
reservation parameter is chosen statically (albeit they can be modified
at run-time by the system administrator) for each core. Once the
reservation parameter is chosen, the primary goal of MemGuard is to
guarantee the reserved bandwidth of each core.
Static partitioning,
however, is inherently inefficient when demands of cores change over time
as unused bandwidth can be wasted.
To minimize bandwidth under-utilization due to the static partitioning,
MemGuard employs a prediction based bandwidth reclaiming mechanism
that dynamically re-distributes unused bandwidths at runtime.

There are, however, a few issues when we apply the MemGuard to
improve performance soft real-time applications. First, MemGuard
reserves memory bandwidth on a per-core basis. Therefore, when a core
hosts different types of applications (with different memory bandwidth
demands), it is difficult to choose appropriate bandwidth reservation
parameters.
Second, while this restriction is mitigated---to a certain degree---by
the runtime bandwidth re-distribution mechanism, the effectiveness of
the mechanism depends on its prediction accuracy, which is in general
very challenging.
In particular, multimedia soft real-time applications, which we focus
on this paper, often show \emph{bursty} memory access patterns, which
are difficult to predict without application level information or a
sophisticated learning algorithm, which is difficult to implement
efficiently in the kernel.

\begin{figure}[t]
\centering
\centering
\subfigure {
\includegraphics[width=0.45\textwidth]{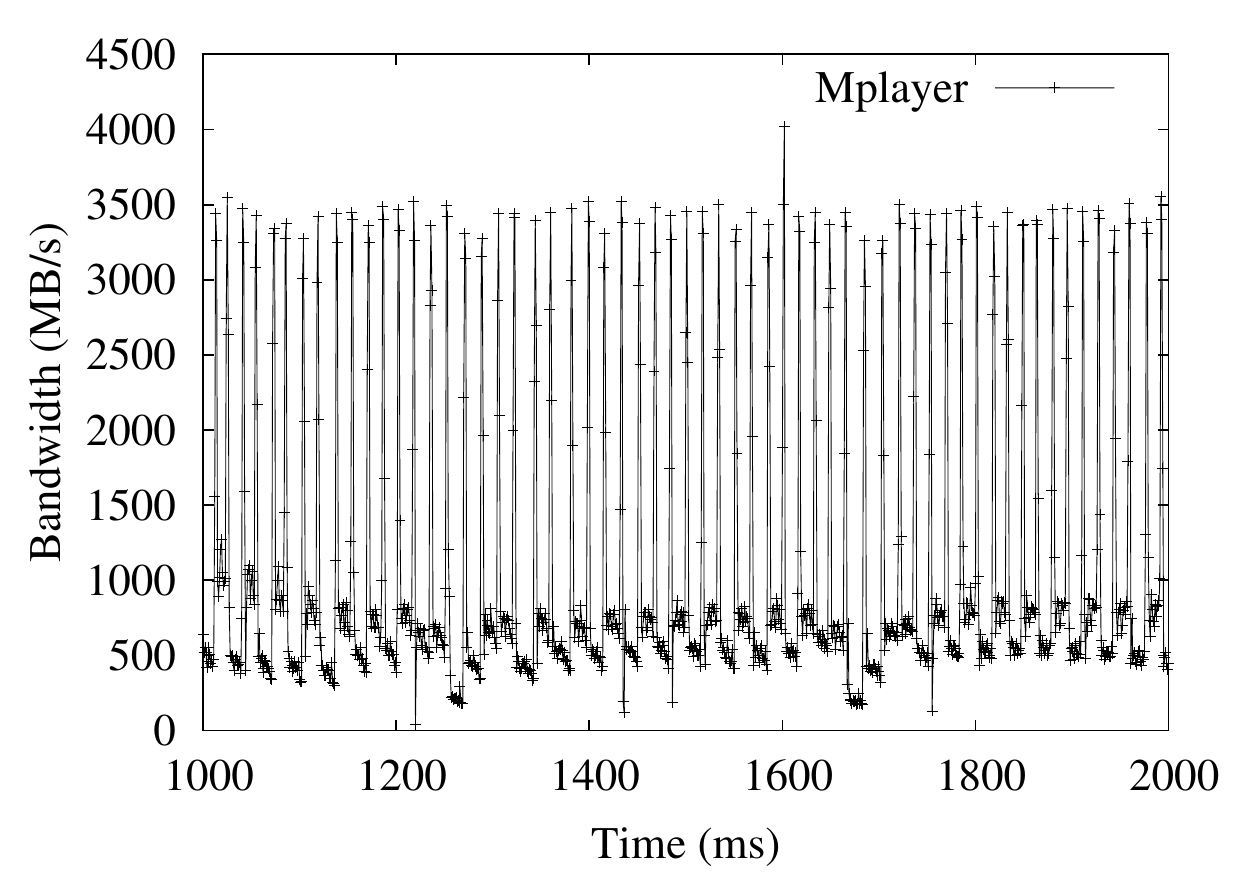}
}
\subfigure {
\includegraphics[width=0.45\textwidth]{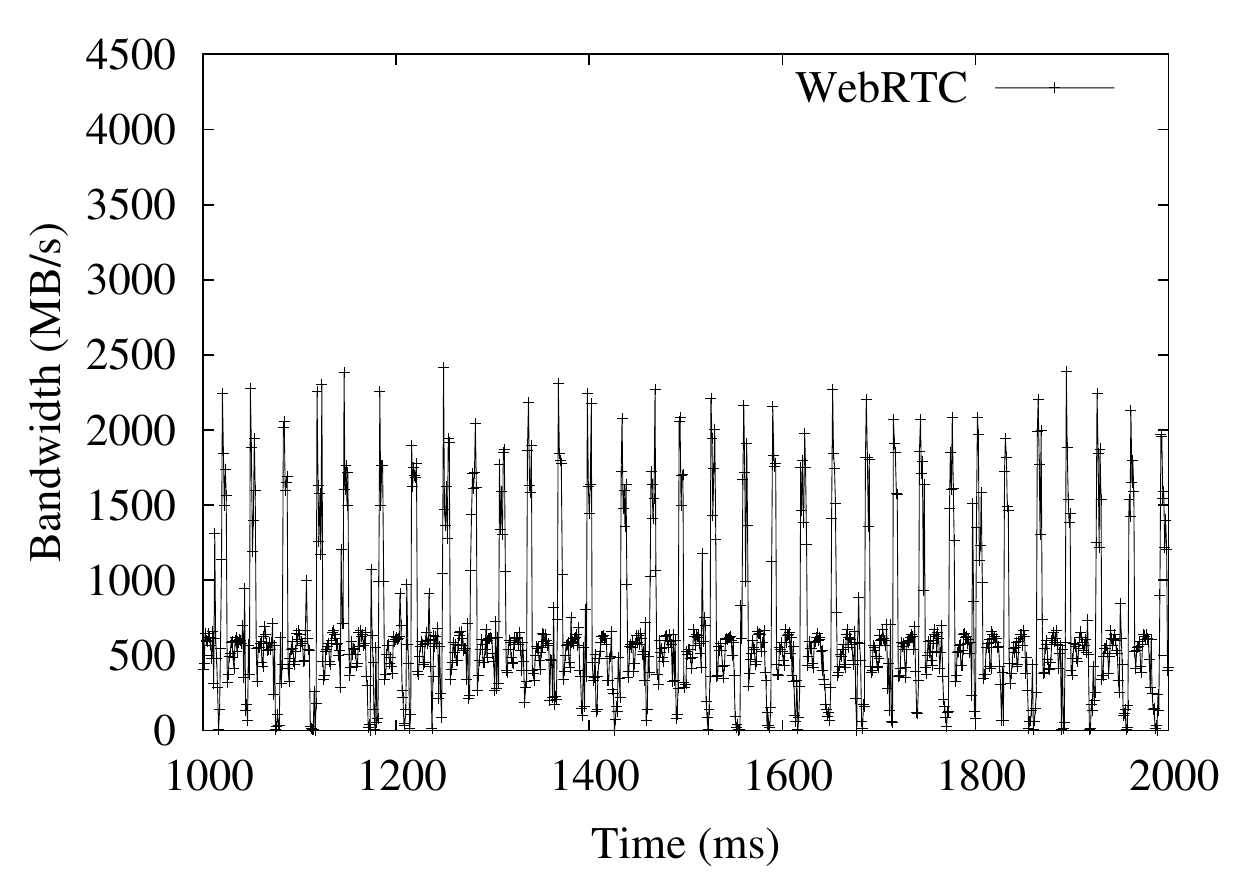}
}
\caption{Memory bandwidth demand changes over time of Mplayer and WebRTC. }
\label{fig:motivation}
\end{figure}


\begin{figure}[t]
\centering
\centering
\includegraphics[width=0.40\textwidth]{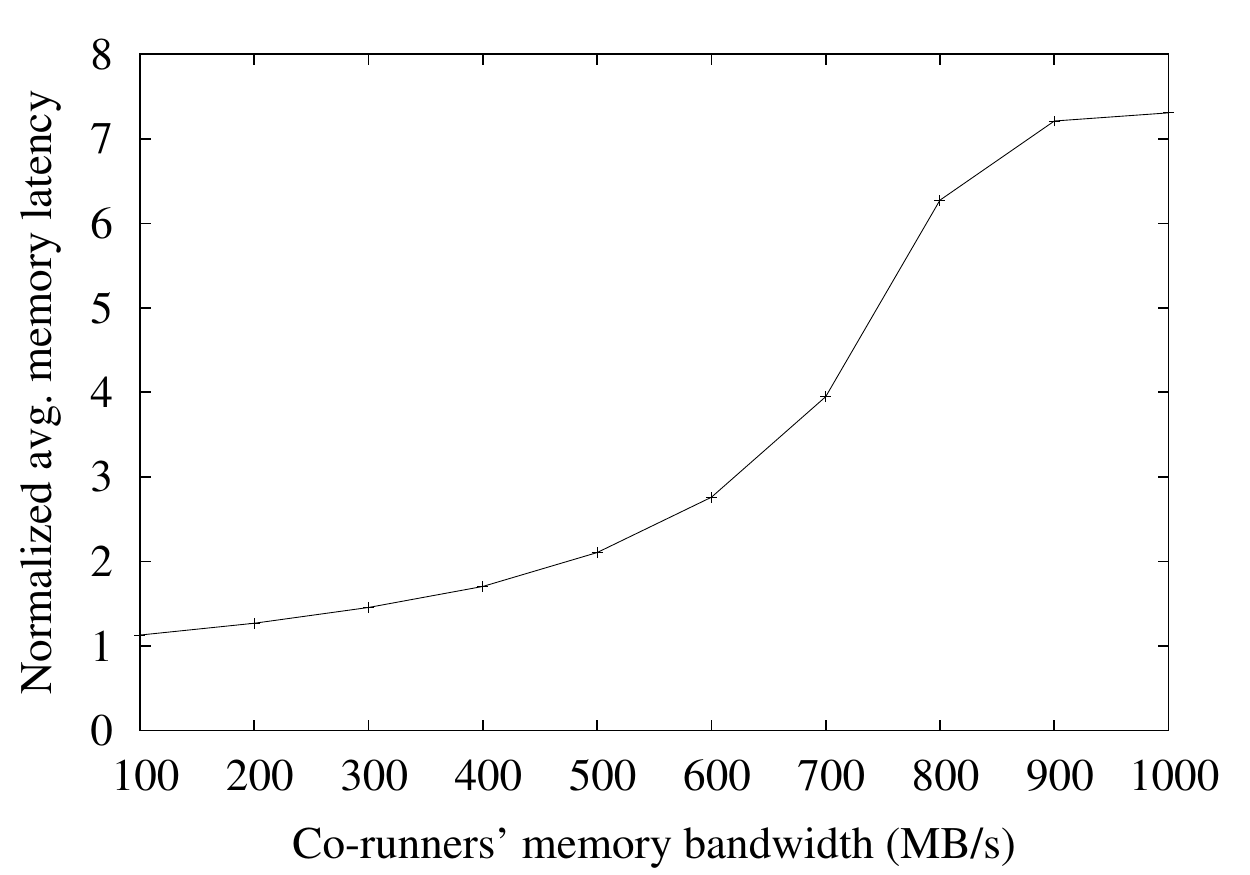}
\caption{Average memory access latency of a Latency
benchmark as a function of the memory bandwidth of
each co-runner on three different cores.}
\label{fig:disproportional}
\end{figure}

Figure~\ref{fig:motivation} shows memory access patterns of two
multimedia applications---Mplayer and WebRTC---collected over a 10
second duration (sampled at every 1ms.)
As both programs process video/audio frames at a regular interval, when
processing a new frame, they require high memory bandwidth for a short
period of time, while at other times their memory bandwidth demands
are low as they are executing compute intensive instructions or
waiting for the next period.

When these soft real-time applications compete memory bandwidth with
other applications running on different cores, the short sections of
code that demand high memory bandwidth could suffer a disproportionally
high degree of performance impact. When the overall memory demand is
low, memory access latencies often can be hidden due to a variety of
latency hiding techniques (e.g., out-of-order) and the abundant memory
level parallelism in modern multicore
architecture~\cite{hennessy2011computer}.
However, when the overall demand is beyond a certain point, such
techniques are no longer able to hide latencies and the requests are
piled up in various queues in the system, which substantially slowdown
all tasks requesting the memory.
Figure~\ref{fig:disproportional} illustrates
this phenomenon. In this experiment, we measure the average memory access
latency (normalized to run-alone performance) of
Latency~\cite{yun2013rtas} benchmark (a pointer chasing
micro-benchmark) while varying the memory bandwidth demand of
co-runners on the other three different cores in a quad-core Intel
Xeon system (Detailed hardware setup is described in
Section~\ref{sec:setup}). When the co-runner's bandwidth is
low (100MB/s), the performance impact to the measured Latency benchmark is
negligible. However, as the memory bandwidth demand of the
co-runners increase (100 - 900 MB/s), the observed delay of
the Latency benchmark is increased exponentially and then saturated
(after 900 MB/s).

In summary, our observations are as follows:
(1) Soft real-time applications such as multimedia applications often
show bursty memory access patterns---regularly requiring a high
memory bandwidth for a short duration of time. (2) Such a period,
which we call a \emph{memory-performance critical section}, is often
critical for timely data processing but it can be disproportionally
delayed by bandwidth demanding co-runners on different cores.

These observations motivate us to design a new memory access control
system, BWLOCK, described in the next section.
\section{BWLOCK} \label{sec:bwlock}

BWLOCK is user-level APIs and a kernel-level memory bandwidth control
mechanism, which is designed to improve performance of soft real-time
time applications (e.g., multimedia applications.) on multicore
systems. It provides simple lock like APIs that can be called
by the applications to express the importance of memory performance
for a given section of code (i.e., memory-performance critical
section.) Once an application acquires the lock, which we call a
memory bandwidth lock, the kernel-level memory bandwidth control
system allows unlimited memory accesses to the requesting task while
regulating the maximum allowed memory bandwidth of the other cores to
avoid excess bandwidth contention, which could delay the task running the
memory-performance critical section.

\subsection{System Architecture}
\begin{figure} [t]
\centering
\centering
\includegraphics[width=0.45\textwidth]{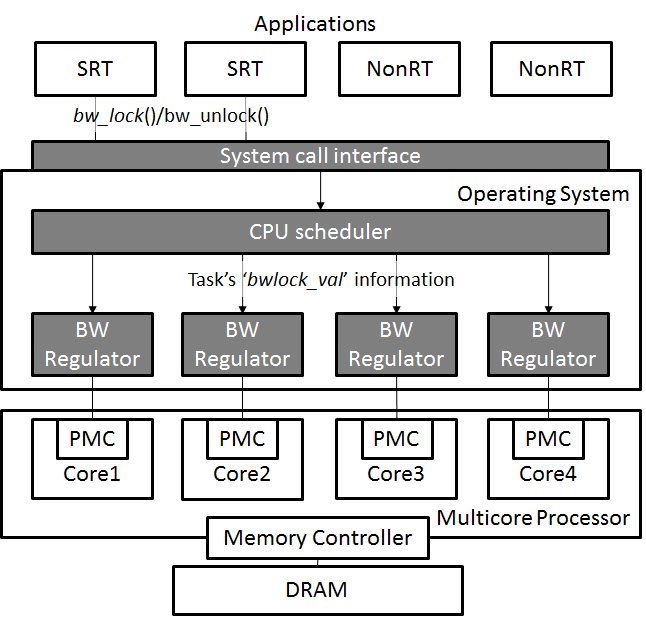}
\caption{Overall system architecture of BWLOCK.}
\label{fig:architecture}
\end{figure}

Figure~\ref{fig:architecture} shows overall architecture of the
proposed system. At the user-level, we provide two
APIs---\texttt{bw\_lock()} and \texttt{bw\_unlock()}---to protect
memory-performance critical sections. When the
\texttt{bw\_(un)lock()} is called, the kernel updates the calling process's
state so that whenever the CPU scheduler schedules the task, the kernel can
determine whether the task is executing the memory critical section or
not. Instead of modifying the code, external utilities can also set
the bandwidth locks of other processes in the system.
The per-core bandwidth regulators are activated when there are one or
more cores executing memory-performance critical sections. In our
current implementation, the check is periodically (e.g., at every 1ms)
performed by software based bandwidth regulators.
Ideally, however, hardware assisted mechanisms could support more
fine-grained memory access control (See Section~\ref{sec:discussion}
for discussions on potential hardware support.)

\subsection{Design and Implementation}

\begin{table}
\centering
\begin{tabular}{c|c}
\toprule
API & Description \\
\midrule
bw\_lock() & begin a memory-performance critical section \\
bw\_unlock() & end a memory-performance critical section \\
\bottomrule
\end{tabular}%
\caption{BWLOCK user-level APIs}
\label{tbl:bwlock-user}
\end{table}

BWLOCK supports fine-grained and coarse-grained bandwidth
locking. In fine-grained mode, programmers are required to use the
APIs in Table~\ref{tbl:bwlock-user} to declare memory-performance
critical sections. It allows fine-grain control over memory
performance but requires detailed profiling information to be
effective. Often, such profiling information can easily be obtained
using publicly available tools such as \emph{perf} in Linux as we will
show in our case studies in Section~\ref{sec:casestudy}. The
coarse-grained mode is an equivalent of calling \texttt{bw\_lock}
once, by the program itself or by the external utility, and never
releases it. Then, whenever the process is scheduled, it automatically
holds the bandwidth lock. We provide an external tool to set the
bandwidth lock of any existing process in the system. Therefore,
BWLOCK can be applied to unmodified programs, albeit the
granularity of control is the entire duration the task occupies a CPU
core. It is important to note that unlike traditional locks used
for synchronization~\cite{block2007flexible}, in which only one task can acquires a lock, a
bandwidth lock can be acquired by multiple tasks, perhaps on different
cores, at any given time. In other words, if there are multiple soft
real-time applications who request a bandwidth lock, all of them will
be granted to access the bandwidth lock. This design is because the
primary goal of BWLOCK is to protect soft real-time applications from
memory intensive non real-time applications. In a sense, our design is
a two-level priority system that prioritizes real-time tasks over non
real-time tasks in accessing memory. It can be, however, naturally
extended to support multiple levels of priorities in accessing memory
in the future. For example, instead of allowing unlimited memory
accesses, the task which holds a bandwidth lock can also be regulated
depending on the priority value associated with the bandwidth lock.


Figure~\ref{fig:bwlock-kernel} shows the kernel-level implementation
of BWLOCK. We added an integer value \texttt{bwlock\_val} to indicate
the status of BWLOCK in the process control block structure of Linux
(\texttt{task\_struct}).
The value can be updated via a system
call (See \texttt{syscall\_bwlock()}). Since it simply updates an
integer value and nothing else, its calling overhead is very small
(Overhead analysis is given in Section~\ref{sec:overhead}).
Each core's bandwidth regulator (See \texttt{per\_core\_period\_handler()})
periodically checks how many cores are executing memory-performance
critical sections (i.e., task's \texttt{bwlock\_val} $>$ 0). If one or
more cores are executing memory-performance critical sections,
only the cores that hold the bandwidth lock can access memory freely
(\texttt{maxperf\_budget} is an infinite value) while the others are
regulated according to \texttt{minperf\_budget}. Note that the
~\texttt{minperf\_budget} is a system parameter that indicates the
maximum amount of memory traffic that can co-exist without significant
performance interference. In our current implementation it is 100MB/s.
If a PMC overflow interrupt occurs, due to exhausting the bandwidth
limit, the core is immediately throttled
by scheduling a high priority real-time kernel thread
(\texttt{kthrottle}). The throttled core is re-activated at the
beginning of every period handler.

\begin{figure}
\begin{lstlisting}
// task structure
struct task_struct {
...
int bwlock_val; // 1 - locked, 0 - unlocked
...
};

// bwlock system call
syscall_bwlock(pid_t pid, int val)
{
struct task_struct *p;
if (pid == 0)
p = current; // 'current' <- calling task
else
p = find_process_by_pid(pid);
p->bwlock_val = val;
return 0;
}

// periodic handler called by the
// bandwidth regulators
void per_core_period_handler()
{
// re-activate the suspended core
if (current == kthrottle)
deschedule(kthrottle);

if (nr_bwlocked_cores() > 0) {
// one or more cores requested bwlock
if (current->bwlock_val > 0)
budget = maxperf_budget;
else
budget = minperf_budget;
} else {
// no cores requested bwlock
budget = maxperf_budget;
}

// program the core's performance counter
// to overflow at 'budget' memory accesses
}

// PMC overflow handler
void per_core_overflow_handler()
{
// stall the core till the next period
// kthrottle <- high priority idle thread
schedule(kthrottle);
}

\end{lstlisting}
\caption{BWLOCK kernel implementation}
\label{fig:bwlock-kernel}
\end{figure}


\section{Evaluation Setup} \label{sec:setup}

In this section, we present details on the hardware platform and
the BWLOCK software implementation. We also provide detailed
overhead analysis and discuss performance trade-offs.

\subsection{Hardware Platform}

We use a quad-core Intel Xeon W3530 based desktop computer as our
testbed. The processor has private 32K-I/32K-D (4/8 way) L1 cache, a
private 256~KiB (8 way) L2 cache for each core and a shared 8MiB (16
way) L3 cache. The memory controller (MC) is integrated in the
processor and connected to a 4GiB 1066~MHz DDR3 memory module. The
graphic card is NVIDIA GeForce 8400.
We disabled turbo-boost, dynamic power management, and hardware
prefetchers for better performance predictability and repeatability.

\subsection{Implementation Details and Overhead Analysis} \label{sec:overhead}

We implemented BWLOCK in Linux version 3.6~\footnote{BWLOCK will be
publicly available at https://github.com/heechul/bwlock}. The kernel's
\texttt{task\_struct} is modified according to
Figure~\ref{fig:bwlock-kernel}. For memory bandwidth control, BWLOCK
uses a modified version of MemGuard kernel module~\cite{yun2013rtas}.

There are two major sources of overhead in BWLOCK: system call and
interrupt handling. First, in the fine-grained setting, two system
calls are required for each memory critical section. In our current
implementation, a single system call is used to
implement both bw\_lock() and bw\_unlock(). The system-call overhead
is small: 125.24ns on average (out of 10,000 iterations.) as it simply
changes a single integer value in the task's \texttt{task\_struct}.

Second, in our current implementation, to monitor which cores
are having the bandwidth lock, a periodic timer handler is being
used as shown in Figure~\ref{fig:bwlock-kernel}. And actual access
control is performed by a performance counter overflow interrupt
handler. Although the overflow handler is not in the critical path of
normal program execution, the period timer interrupt is pure overhead
that is added to the task's execution time, just like the OS tick
timer handler in standard operating systems. We quantified
the period interrupt handling overhead by measuring the execution
increases of a benchmark. Table~\ref{tbl:overhead-period} shows the
measured overhead (i.e., percentage of the increased execution time.)
under different period lengths. Based on this result, we use 1ms
period unless noted otherwise.

\begin{table}
\centering
\begin{tabular}{rl}
\toprule
Period (us) & Overhead (\%) \\
\midrule
100 & 3.5 \\
250 & 1.5 \\
500 & 0.9 \\
1000 & 0.7 \\
2500 & 0.5 \\
\bottomrule
\end{tabular}
\caption{Period interrupts handling overhead}
\label{tbl:overhead-period}
\end{table}

\section{Evaluation Results} \label{sec:casestudy}

In this section, we presents case-study results using two real-world
soft real-time applications---Mplayer (a video player) and
WebRTC~\cite{webrtc} (a multimedia real-time communication framework
for browser based web applications)---to evaluate the effectiveness
of BWLOCK.

\subsection{Mplayer}

Mplayer is a widely used open-source video player. In the following
set of experiments, our goal is to protect real-time performance of the
Mplayer(s) in the presence of memory intensive co-running applications
while still maximizing overall throughput of the co-runners.

In the first set of experiments, one Mplayer instance plays an H264
movie clip with a frame resolution of 1920$\times$816 and a frame rate
of 24fps. We slightly modified the source code of Mplayer to get the
per-frame processing time and other statistics.
Decoded video frames are displayed on screen via a standard
X11 server process. Therefore, the Mplayer and the X11 have soft
real-time characteristics.

\subsubsection{Profiling}
To understand their memory-performance characteristics, we
collect function level profiling information---cache-misses and cycles
of each function---with the \emph{perf} tool, which uses hardware
performance counters. The profiled information of Mplayer and X11 is
shown in Table~\ref{tbl:mplayer-profile} and \ref{tbl:x11-profile},
respectively. In both cases, the functions that generate most of
memory traffic were identified: \texttt{yuv420\_rgb32\_MMX} in Mplayer
and \texttt{sse2\_blt} in X11. Note that each function is responsible
for more than 50\% of total LLC-misses of its respective application,
while is responsible for much less CPU cycles (27.8\% and 32.85\%
respectively). Therefore, they are prime candidates for applying
BWLOCK. Due to the restrictions of our current
implementation---integration of periodic bandwidth regulation---it
is also important to know the duration of each function: if it is too
short, BWLOCK may not be able to regulate co-runners' memory
accesses when needed. Table~\ref{tbl:functions-time} shows the
average and 99 percentile execution times of the
functions. Fortunately, both functions are long enough to be regulated
by the bandwidth control mechanism of BWLOCK.

\begin{table}[htbp]
\centering
\begin{tabular}{rrl}
\toprule
LLC misses & Cycles & Function \\
\midrule
51.6\% & 27.8\% & yuv420\_rgb32\_MMX \\
18.8\% & 9.3\% & prefetch\_mmx2 \\
4.5\% & 7.3\% & hl\_decode\_mb\_simple\_8 \\
\bottomrule
\end{tabular}%
\caption{Profiled information of Mplayer}
\label{tbl:mplayer-profile}%
\end{table}%

\begin{table}[htbp]
\centering
\begin{tabular}{rrl}
\toprule
LLC misses & Cycles & Function \\
\midrule
53.29\% & 32.85\% & sse2\_blt\\
24.13\% & 24.19\% & fbBlt\\
14.10\% & 19.61\% & sse2\_composite\_over\_8888\_88888 \\
\bottomrule
\end{tabular}%
\caption{Profiled information of X11}
\label{tbl:x11-profile}%
\end{table}%

\begin{table}[htbp]
\centering
\begin{tabular}{rrll}
\toprule
Average & 99 pct. & \multirow{2}[0]{*}{Function} & \multirow{2}[0]{*}{Application} \\
duration & duration & & \\
\midrule
2.9ms & 4.2ms & yuv420\_rgb32\_MMX & Mplayer \\
1.1ms & 2.9ms & sse2\_blt & X11 \\
\bottomrule
\end{tabular}%
\caption{Timing statistics of memory intensive functions}
\label{tbl:functions-time}%
\end{table}%



\subsubsection{Performance comparison}
To investigate the effectiveness of BWLOCK, we conducted a set of
experiments. We first run the Mplayer alone (with the
X-server) to get the baseline performance. In order to generate memory
interference, we use two instances of a memory intensive synthetic
benchmark~\cite{yun2013rtas}, referred as \emph{bw\_write}. We also
measure their baseline performance in isolation. We co-schedule all
four processes---Mplayer, X11, and two bw\_write instances---at the
same time in four different configurations.
For convenience of monitoring and measurements, each process is
assigned to a dedicated core using a $cpu$affinity facility in
Linux. Note that all four processes are single-threaded.
In \emph{Default}, we use a standard vanilla Linux kernel. In
\emph{MemGuard}, we use MemGuard~\cite{yun2013rtas}; the memory bandwidth
budgets are configured as 450, 450, 100, and 100MB/s for Core0 to 3,
respectively, and predictive bandwidth re-distribution is enabled;
note that Mplayer (Core0) and X11 (Core1) are reserved more bandwidths
than the co-runners. In \emph{BWLOCK(fine)}, we manually insert
\texttt{bw\_lock} and \texttt{bw\_unlock} in the previously identified
memory intensive functions of Mplayer and X11 (Table~\ref{tbl:functions-time}), as shown in
Figure~\ref{fig:bwlock-example}. Lastly, in \emph{Bwlock(coarse)},
both Mplayer and X11 are \emph{not} modified but configured to automatically
hold the bandwidth lock whenever they are scheduled.

\begin{figure}
\begin{lstlisting}
static inline int yuv420_rgb32_MMX
(SwsContext *c, const uint8_t *src[],
..
{
bw_lock(); // added

YUV2RGB_LOOP(4)
...
bw_unlock(); // added
}
\end{lstlisting}
\caption{Code modification example for fine-grained application of BWLOCK. }
\label{fig:bwlock-example}
\end{figure}


\begin{figure}
\centering
\includegraphics[width=0.5\textwidth]{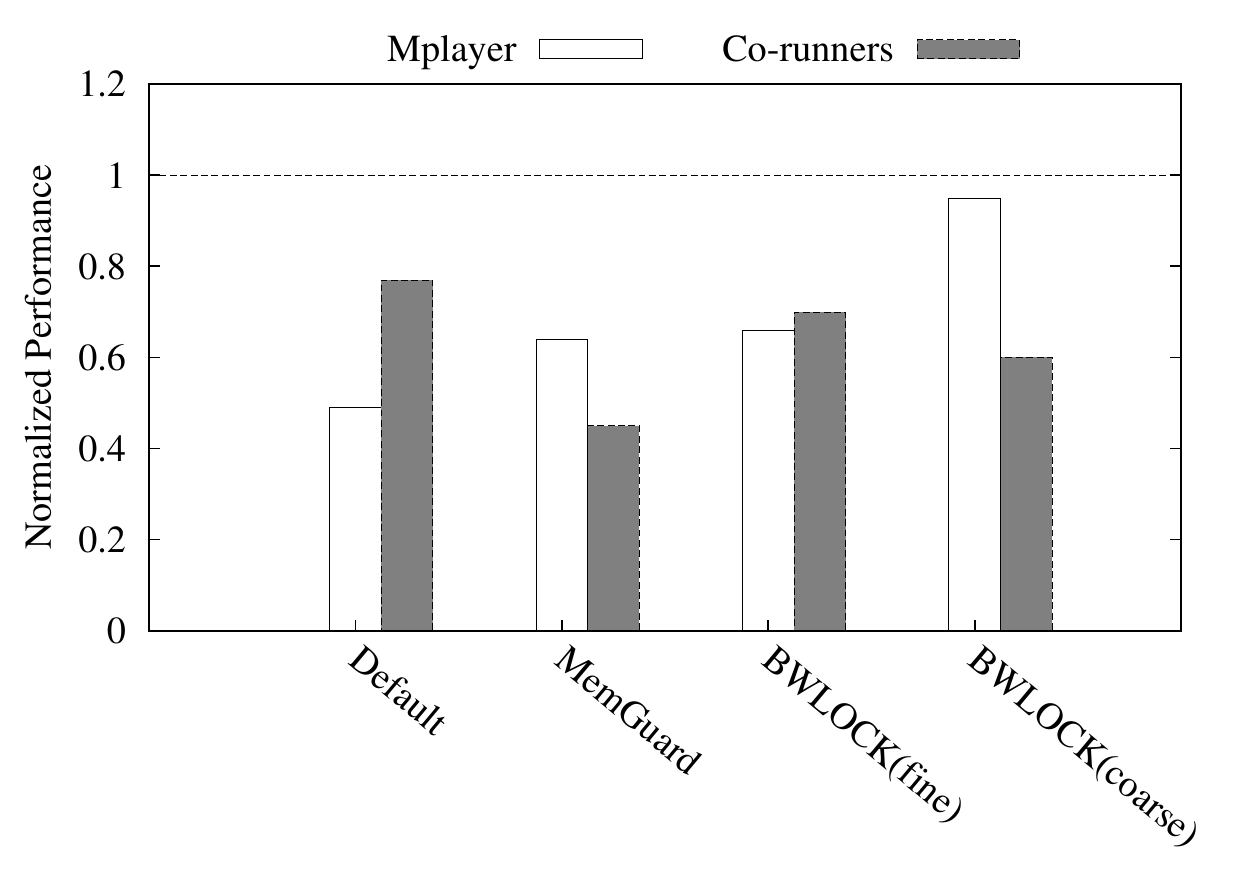}
\caption{Normalized performance of Mplayer (average frame time) and
co-running Bandwidth benchmarks (MB/s): 1$\times$Mpalyer and 2$\times$Bandwidth instances.}
\label{fig:result-mplayer}
\end{figure}

\begin{figure*}
\centering
\centering
\subfigure [Default] {
\includegraphics[width=0.45\textwidth]{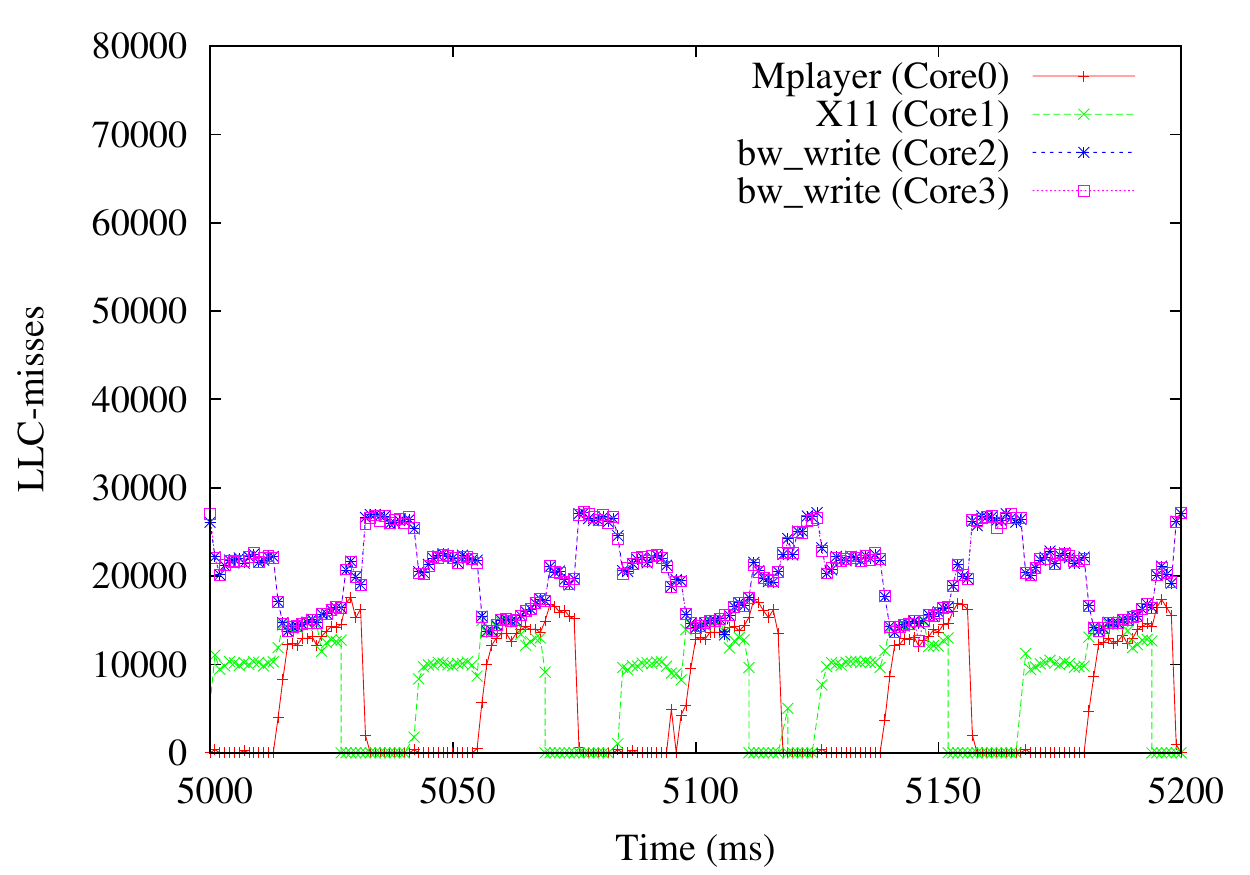}
}
\subfigure [MemGuard] {
\includegraphics[width=0.45\textwidth]{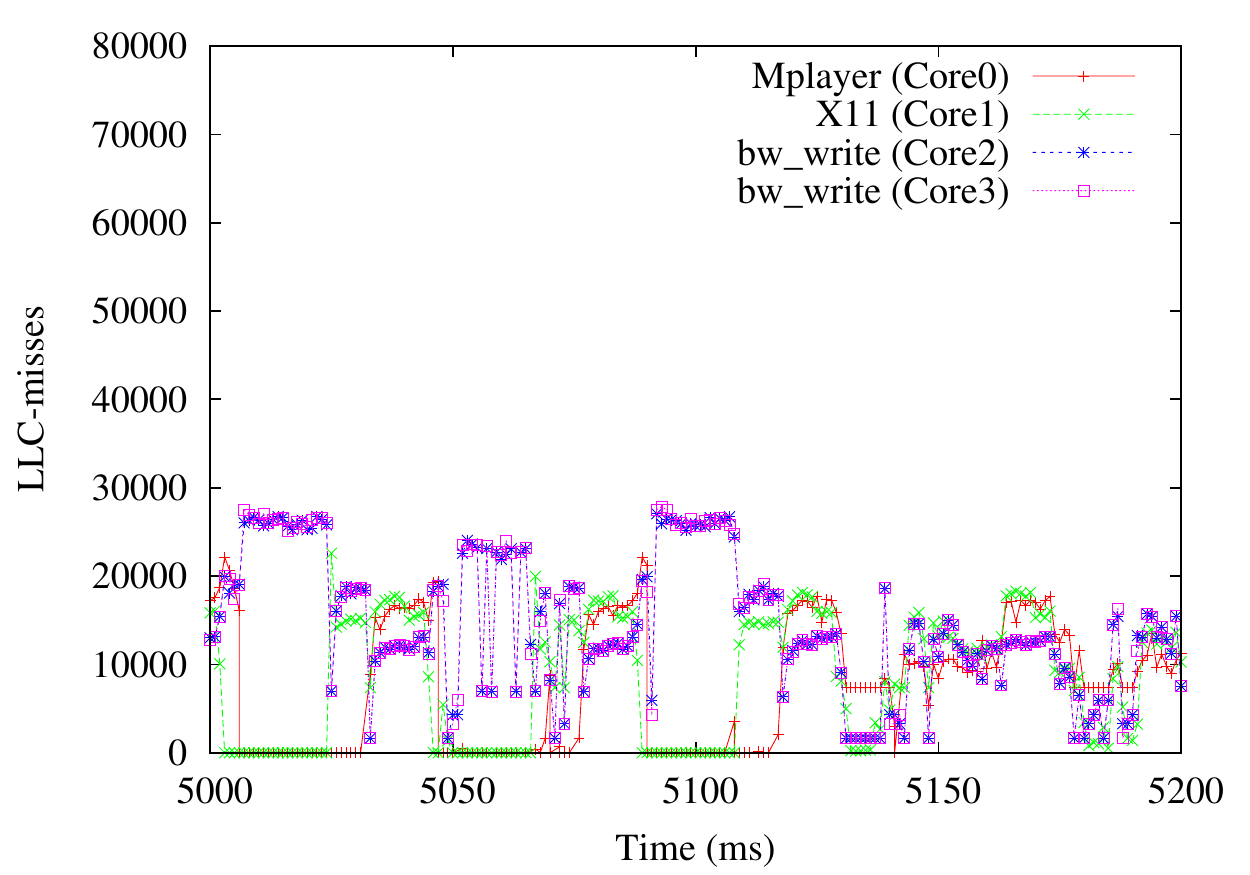}
}
\subfigure [BWLOCK(fine)]{
\includegraphics[width=0.45\textwidth]{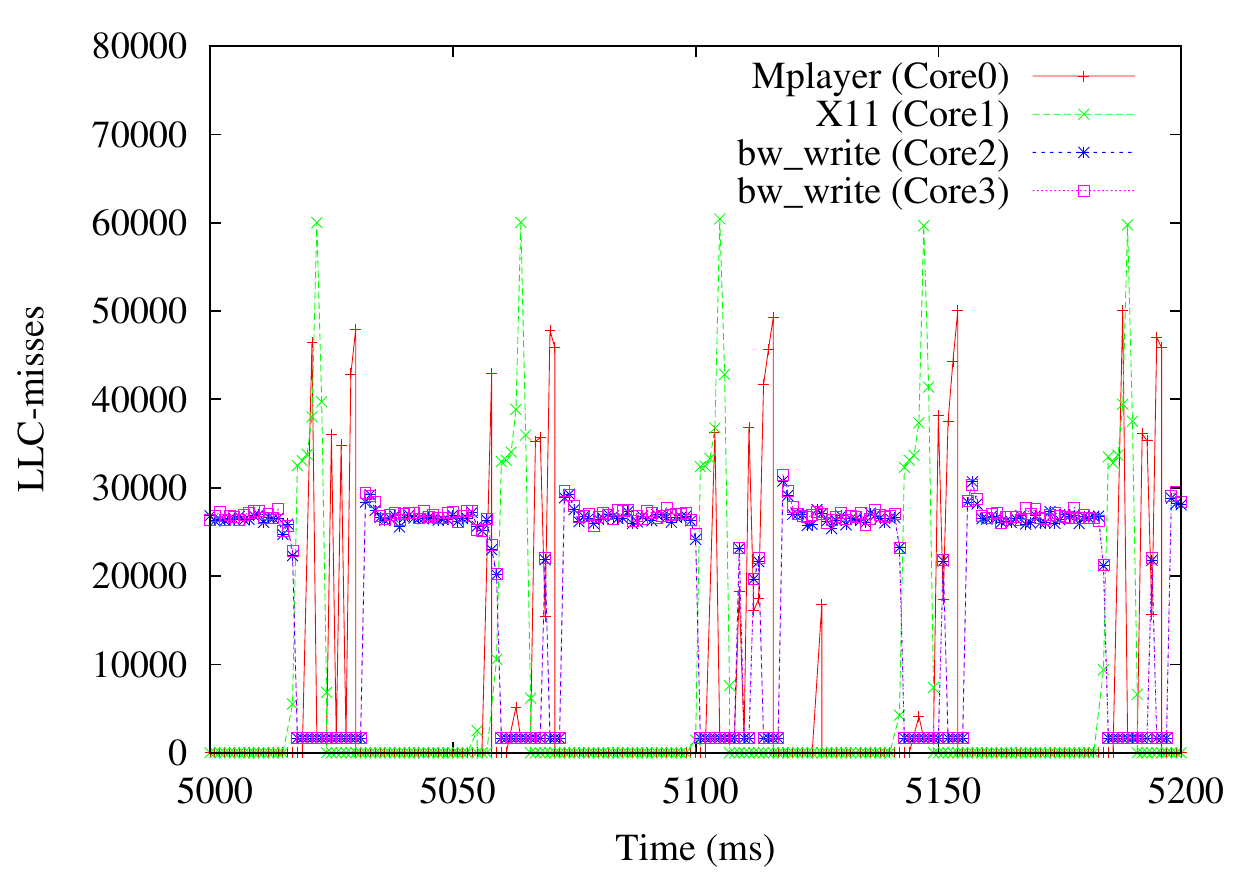}
}
\subfigure [BWLOCK(coarse)]{
\includegraphics[width=0.45\textwidth]{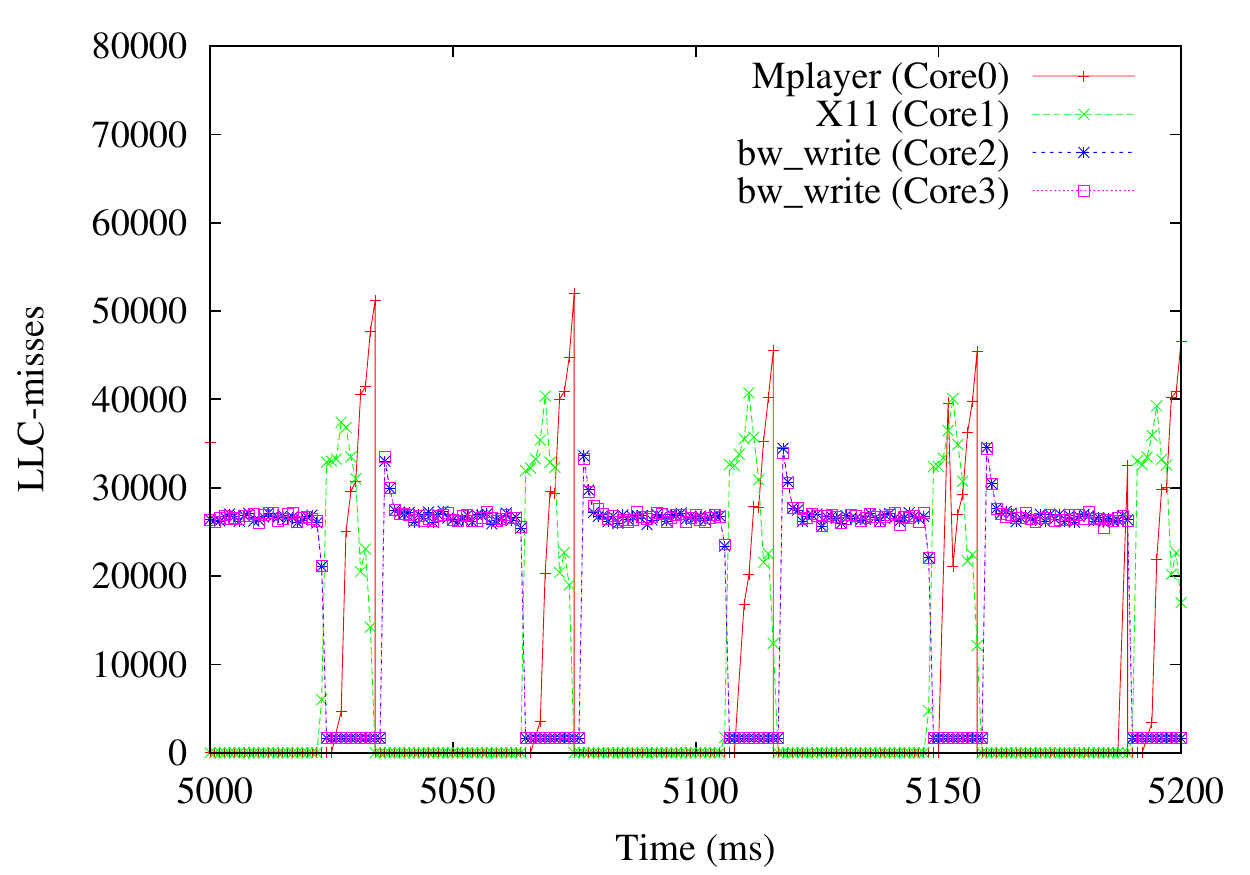}
}
\caption{Per-core memory access patterns. }
\label{fig:percore-view}
\end{figure*}

Figure~\ref{fig:result-mplayer} shows results. For Mplayer,
performance is measured by the average frame processing time,
normalized to run-alone performance. For co-running \emph{bw\_write}
benchmarks, performance is measured by the aggregated throughput
(MB/s) of the two. In the figure, performance is normalized to each
application's baseline performance measured in isolation. In Default,
Mplayer's performance is significantly suffered---dropped by 51\%---due to
memory contention with the co-running bw\_write instances, which are
much less affected---dropped by 22\%. This kind of disproportional
performance impact is common in COTS multicore systems and is caused
by a combination of application memory characteristics and DRAM
controller's scheduling
policy~\cite{moscibroda2007memory,kim2010thread}.
In MemGuard, Mplayer's performance is better protected---dropped by
32\%---as more memory bandwidth is reserved for it. However, this
comes at a cost of considerable performance reduction of the
co-runners---only 51\% the baseline performance. In BWLOCK(fine), on
the other hand, both Mplayer's and co-runners' performance are
improved over MemGuard---by 12\% for Mplayer and 13\% for
co-runners. This is because memory-performance critical sections in
the Mplayer, identified from profiling, are protected from being
interfered by the co-runners' memory accesses using the explicit
\texttt{bw\_lock} and \texttt{bw\_unlock}. Lastly, in BWLOCK(coarse),
the Mplayer is unmodified but whenever it is scheduled, it
automatically calls the bandwidth lock by the CPU scheduler. As a
result, the Mplayer's performance is almost identical to the baseline
performance. However, because the entire duration of Mplayer's
processing is protected by the bandwidth lock, even if it doesn't
access memory, the co-runners' performance is slightly further
degraded.

Figure~\ref{fig:percore-view} shows the memory access pattern of each
core. The y-axis shows the number of LLC misses of each core for every one
millisecond period. Note that Core2 and Core3 have a constant memory
demand when they run in isolation. In Default, whenever Mplayer and/or
X11 begin processing and demand high memory bandwidth, all tasks
suffer considerable bandwidth contention. In MemGuard, we can observe
that Mplayer (and X11) is getting more bandwidth than the
bw\_write when needed. However, due to difficulties of making accurate
predictions on future usage, which MemGuard relies on, its demand is
not always satisfied. In both BWLOCK(fine) and BWLOCK(coarse), on the
other hand, we can observe co-runners are immediately regulated upon
arrivals of Mplayer's memory demands; hence Mplayer can achieve near
identical to its baseline performance in isolation.

\begin{figure}
\centering
\centering
\includegraphics[width=0.45\textwidth]{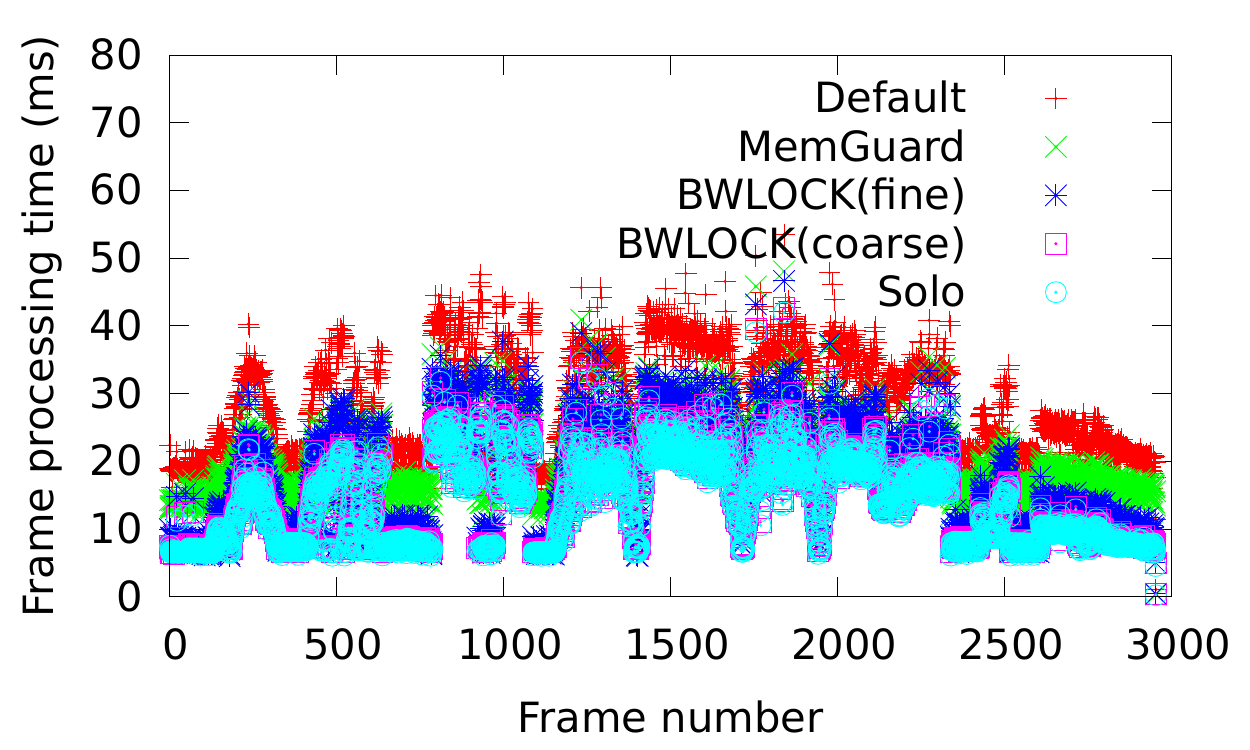}
\caption{Frame processing time comparison.}
\label{fig:frameprocessing}
\end{figure}

Figure~\ref{fig:frameprocessing} shows frame processing time in
different system configurations. Note that Solo represents Mplayer's
baseline performance measured in isolation. BWLOCK(coarse) is mostly
overlapped with Solo. BWLOCK(fine) and MemGuard take longer in
processing frames and Default, as expected, takes the longest in most
frames.

\subsubsection{Overloaded System}
\begin{figure}
\centering
\includegraphics[width=0.45\textwidth]{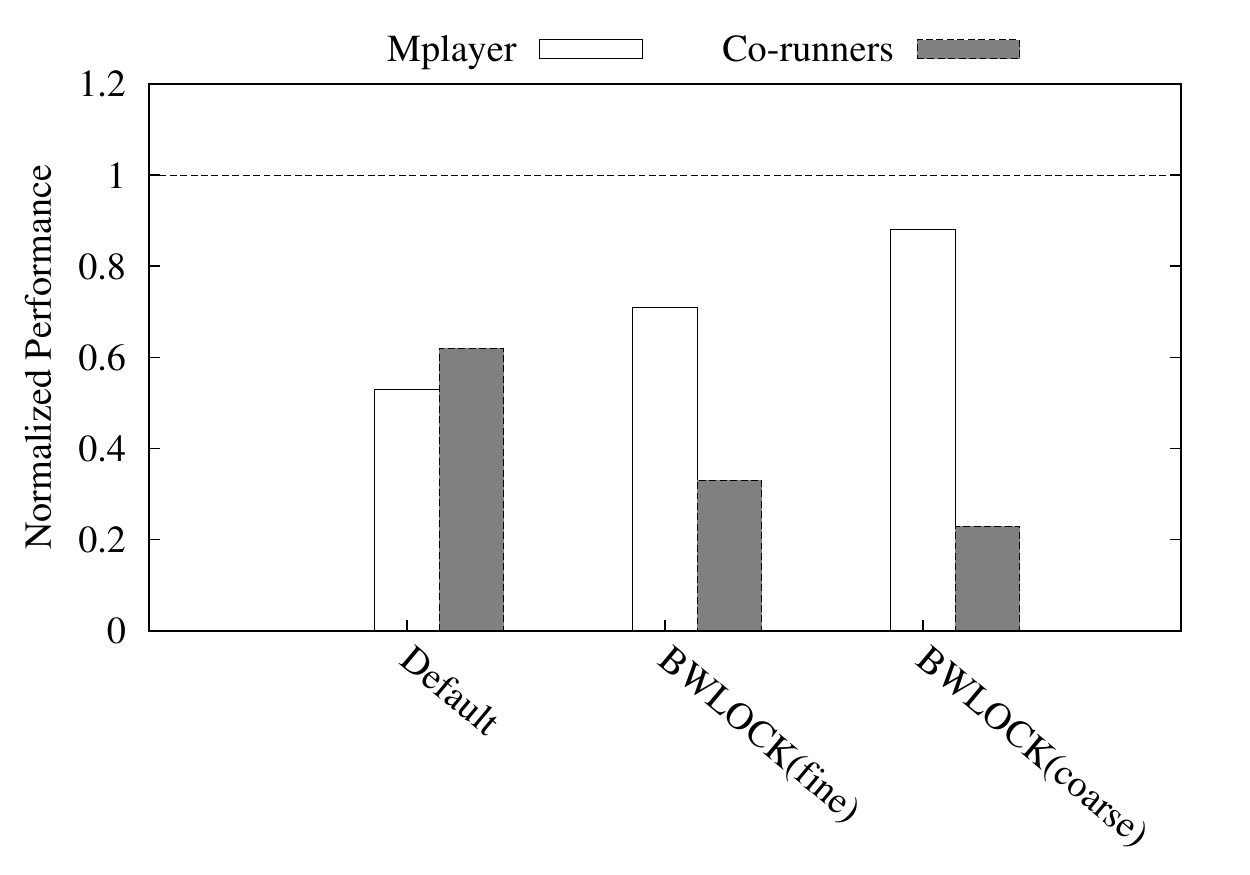}
\caption{Normalized performance of Mplayer (average frame time) and
co-running Bandwidth benchmarks (MB/s): 4$\times$Mplayer and 4$\times$Bandwidth instances.}
\label{fig:result-mplayer-4bw4mp}
\end{figure}

So far, we have assigned one task per core and both Mplayer and X11 do not
consume 100\% cycles of the assigned core. In other words, the system
is under-utilized. In order to investigate how BWLOCK performs in an
overloaded system, we performed another set of experiments in which
each core runs a Mplayer and a bw\_write instance (i.e., four Mplayer
instances and four bw\_write instances) to fully load the
system. Performance metrics are the same: average frame processing
time of Mplayer and the aggregate bandwidth of
bw\_write. Figure~\ref{fig:result-mplayer-4bw4mp} shows the
results. Notice that, in this experiment setup, all cores run both
real-time and non-real-time tasks. Therefore, MemGuard's core-based
bandwidth partitioning, which prioritizes certain cores over the
others, is not appropriate. Hence, we only compare the results of
Default and the two BWLOCK settings (fine and coarse). As shown in the
figure, both BWLOCK settings provide good performance isolations for
the Mplayer instances at the cost of more degraded performance for the
co-runners which do not request bandwidth locks. Note that our current
BWLOCK implementation does not limit the number of tasks that can hold
bandwidth locks at a given time. Therefore, memory contention among
the soft real-time tasks, which hold bandwidth locks on different
cores, could potentially cause delay with each other. The performance
reduction of Mplayer in BWLOCK(coarse) is not from the contention from
the bw\_write instances but is entirely from the co-running Mplayer
instances---we verified this by comparing it with the result obtained
by running only four instances of Mplayer without the bw\_write
instances.

\subsection{WebRTC}
WebRTC is an open source, plug-in free, RTC (real-time communication)
platform for enabling audiovisual, network-based applications between
browsers. The goal of this experiment is to provide real-time
performance isolation to WebRTC sessions, in the presence of memory
intensive co running applications on multi-core platforms. We also
investigate the side effects of different isolation mechanisms on the
performance of co runners. The setup is configured to achieve
negligible congestion in the network by having two communication hosts
directly connected through a Gigabit Ethernet switch. Hence, the
performance variability observed is entirely because of resource
contention in the host itself. WebRTC utilizes GCC (Google Congestion
Control) algorithm to derive target bit-rate of audiovisual streams
based on the resource contention in network, and the end hosts
\cite{ccm2013webrtc}. The frame rate and sending bandwidth are
adjusted to match the available resources at any given time. The
default resolution of 640$\times$480, and frame rate of 30 FPS is used
for experimentation, while the threshold bandwidth is increased to 4
Mb/s from default 2 Mb/s. LBM benchmarks from SPEC2006 suite are
chosen as co-running applications. Since X11 server is the front end
of the WebRTC, they are considered together as group, and assigned to
share the CPU cores in cgroups. While, lbm co-runners are allocated to
remaining two CPUs belonging to another CGROUP.
\subsubsection{Profiling}
Similar to MPlayer, to understand the memory access pattern, we
collected function level profiling information for WebRTC, using Linux
\emph{perf} tool. This time we only focused on \emph{cache-miss}
events, to understand the memory access behavior. Functions
\texttt{sk\_memset32\_SSE2} and \texttt{S32A\_Opaque\_BlitRow32\_SSE2}
from Skia library seems to cause more than 50\% (29.29\% and 22.95\%
respectively) of cache misses during a WebRTC sessions. The mean
execution length of these functions is 7.5 us, while more than 99\% of
sample values being less than 100 us. The function execution length
is much smaller than that were observed with MPlayer profiled
functions. These functions didn't seem ideal for applying fine grained
BWLOCK, as the minimum BWLOCK period is 1 ms. We think that large
number of invocation of these low level graphics functions are being
made by higher level subroutine(s). Bursty invocation of these
functions might lead to aggregated continuous time periods (during
which these functions are active), in the order of regulation period
of BWLOCK. So, we experimented fine grained BWLOCK on above two
discovered functions to understand the effects of fine grained memory
bandwidth regulation, compared it's performance with other isolation
techniques, namely, coarse-grain BWLOCK, MemGuard, and Default (CPUSET
partitioning). Similar to MPlayer approach, entry and exit
(\texttt{bw\_lock} and \texttt{bw\_unlock}) calls are introduced
during which sufficient memory bandwidth (1000 MB/s) is reserved for
the corresponding cpu cores, while other cores bandwidth quota is set
to 100 MB/s.

\begin{figure}
\centering
\includegraphics[width=0.5\textwidth]{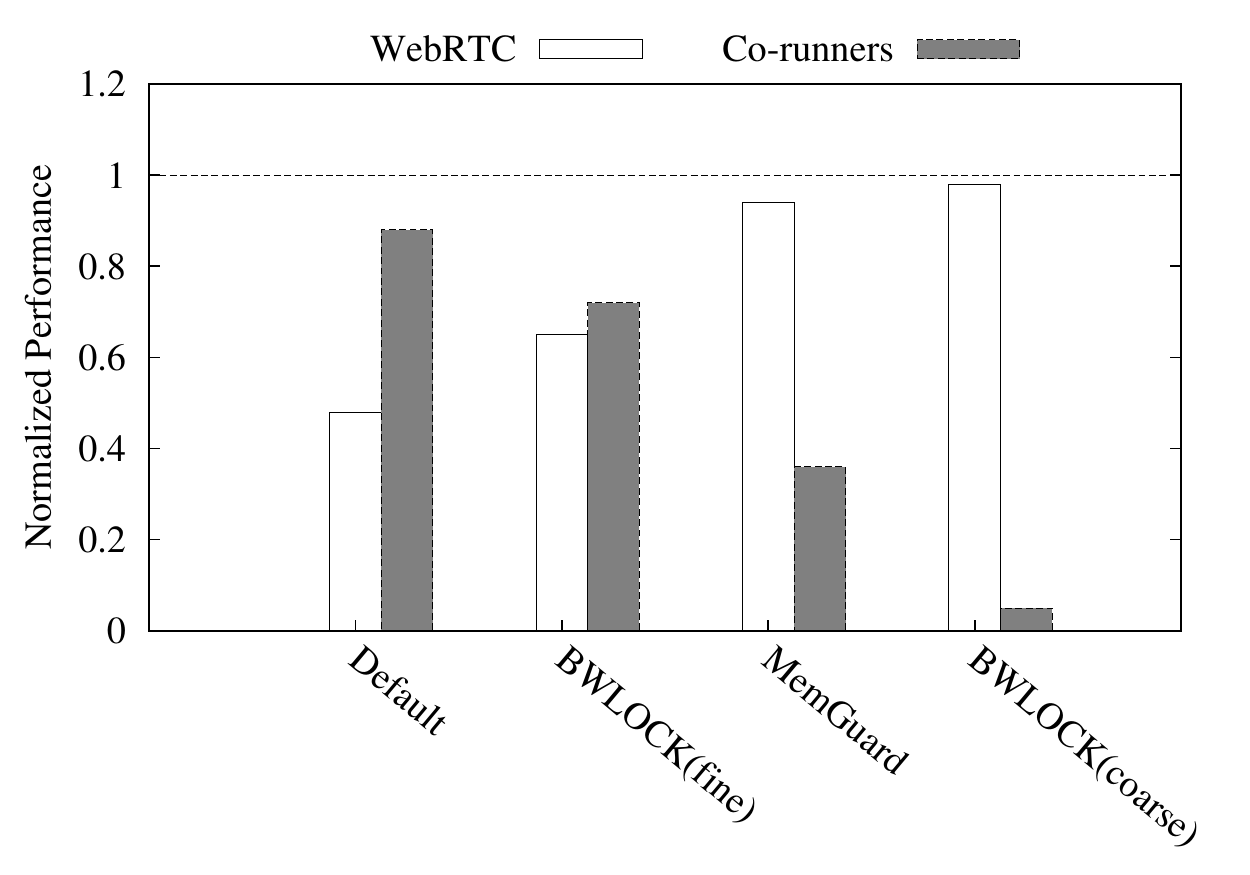}
\caption{Normalized performance of WebRTC (average bandwidth) and
co-running LBM benchmarks (MB/s)}
\label{fig:result-webrtc}
\end{figure}

\subsubsection{Performance}
Figure \ref{fig:result-webrtc} shows the normalized performance of
WebRTC and co-running LBM(s) with different isolation mechanisms.
Coarse-grain BWLOCK achieves near perfect performance isolation for
WebRTC from co running LBM tasks, albeit with heavy penalization for
co runners. WebRTC process consists of \url{~}20 threads, and out of which,
couple of threads are involved in encoding and decoding of
video. Hence, the coarse grain mechanism over reserves the bandwidth
for WebRTC process, leaving very small spare bandwidth for co
runners. With fine grained BWLOCK (by using \texttt{bw\_lock} and
\texttt{bw\_unlock}) on profiled graphic functions, the performance of
WebRTC improves without much penalty to co running tasks. Since the
two profiled functions contribute around 50\% of cache misses, perfect
isolation is not achieved, at the same time, many non-core threads
(threads not involved in encoding and decoding of video) are not
bandwidth reserved leaving sufficient room for LBM co runners. Some
performance penalty is incurred due to very small execution duration
of profiled function leading to incensed overhead of system calls. By
using MemGuard in reclaim and sparing sharing mode, we could achieve
perfect isolation for WebRTC performance with more 50 \% penalty for
co runners. In comparison to MemGuard, BWLOCK is a dynamic, on-demand
kind of mechanism, whereas, MemGuard requires static, pre-determined,
per core bandwidth allocation. All the approaches achieve better
real-time WebRTC performance compared to Default (CPUSET partitioning
alone).\\
Table \ref{tbl:webrtc-metrics} shows the important metrics of
WebRTC. The results correspond to the average bandwidth achieved by
WebRTC in specific scenarios. Except for Default (CPSET alone
partitioning) and fine-grained BWLOCK, all configurations provide
complete isolation to WebRTC from co-runners, albeit, having varying
degree of penalty on co running applications. A clear trade-off
emerges, with Default, BWLOCK(fine), MemGuard, and BWLOCK(coarse)
providing increasing levels of isolation to WebRTC, while increasing
penalty for co runners. As GCC kicks in during resource contention,
the bandwidth/frame rate is dynamically adapted leading to reduction
in bandwidth and/or frame rate. These parameters together determine
the achieved audiovisual quality.

\begin{table}[htbp]
\centering
\begin{tabular}{lccc}
\toprule
Config. & RTT (ms) & FR (FPS) & BW (kb/s) \\
\midrule
Default & 17.20 & 21.34 & 2917.85 \\
BWLOCK(fine) & 4.22 & 29.58 & 2229.10 \\
MemGuard & 2.24 & 29.98 & 4019.16 \\
BWLOCK(coarse) & 2.22 & 30.00 & 4025.30 \\
\bottomrule
\end{tabular}%
\caption{WebRTC internal performance metrics}
\label{tbl:webrtc-metrics}%
\end{table}%

\section{Discussion} \label{sec:discussion}
In this section, we discuss limitations of our approach and future
improvements. 

\subsection{Hardware Assisted Memory Bandwidth Control}
A significant limitation of our current approach is our software based
periodic monitoring and bandwidth controlling mechanism in which the
control granularity is limited to a millisecond range due to the
interrupt handling overhead. This means the detection and application
of bandwidth lock can be delayed up to the timer period. While this
may not be a serious issue in many soft real-time applications as we
have shown in this paper, there may be other applications in which such
delay are not tolerated. This limitation can easily be overcome via
hardware support in the memory controller or the CPU. For example, 
hardware can expose a set of registers---that control the memory
access priorities in the DRAM controller~\cite{kim2010thread} or the
size of MSHR in the shared cache~\cite{ebrahimi2010fairness}---to the
kernel. Then BWLOCK can simply update such registers to protect memory
performance critical sections. 

\subsection{Application to Hard Real-Time Systems}
Although we focus on soft real-time applications, we believe BWLOCK can
also be applied to hard real-time systems in some cases. For example,
it is possible to designate a single core to execute all hard
real-time applications while the other cores execute non real-time
applications. In such a scenario, we can apply BWLOCK to all hard
real-time tasks on the designated core to ensure that while any of the
hard real-time tasks execute, all other cores' maximum memory
bandwidth usage could be limited to a certain number. Then, non
real-time tasks and hard real-time tasks can safely co-exist without
needing to worry about excessive memory contention. Especially, with
hardware support mention earlier, such design can be used for systems
that need certification~\cite{faa2014certification}. 
\section{Related Work} \label{sec:related}

OS level memory access control was first discussed in literature by
Bellosa \cite{bellosa97processcruisecontrol}. The basic idea is to 
reserve a fraction of memory bandwidth for each
core~\cite{bellosa97processcruisecontrol,yun2012ecrts,yun2013rtas} (or
task~\cite{inam2014multi}) by means of software mechanisms---e.g., TLB 
handler~\cite{bellosa97processcruisecontrol} or hardware performance
counter interrupts~\cite{yun2013rtas,inam2014multi}. One problem of
the memory bandwidth reservation approach is that by partitioning
memory bandwidth among the cores (or tasks), usable bandwidth can be
substantially wasted if the reserved bandwidth is not being fully used
by the reserved core (task). The work in \cite{yun2013rtas} partly solves the
problem by supporting dynamic reclaiming and sharing that re-distribute
memory bandwidth of the cores that under-utilize their reserved
bandwidth to the cores that need more than their reserved bandwidth.

However, the effectiveness of the techniques depends on cores' memory access
patterns and the accuracy of future usage predictions. 
In general, memory bandwidth reservation systems are not ideal in
efficiently utilizing available memory bandwidth---which is essential in
many soft real-time systems where real-time applications are co-scheduled
with non real-time applications---because the reserved bandwidth for
certain real-time tasks would result in under-utilization of the
memory subsystem. In contrast, BWLOCK allow unrestricted memory 
accesses for most of the time, hence leveraging full benefits of
parallelism available in modern multicore architecture, but limit
excessive concurrent memory accesses from non real-time tasks only
when doing so would likely affect performance of the soft real-time
tasks that are executing memory-performance critical sections. 
We find that these selective regulations are more efficient in 
utilizing memory bandwidth while still providing good real-time
performance than the reservation based approaches.

In the context of proving performance isolation in multicore systems,
software based cache partitioning technique, known as page coloring,
has been extensively studied~\cite{lin2008gaining,zhang2009towards,ding2011srm,kim2013coordinated,mancuso2013rtas,ye2014coloris,yun2014rtas}. The
basic idea is to allocate memory pages of certain physical addresses
such that each core accesses different part of the cache-sets. This
way, cache can be effectively partitioned without needing special
cache hardware. A downside of this approach is, however, that it is
very costly to change the size of partition at runtime. 
More recently, page coloring has been applied to partition DRAM
banks~\cite{liu2012software,suzuki2013coordinated,yun2014rtas}. In
line with the problems of bandwidth partitioning, however, these
shared space resource partitioned (cache and DRAM bank space)
resources can be wasted if they are not utilized by the reserved cores
or tasks. Nevertheless, these space partitioning techniques can
reduce the degree of interference experienced by concurrent tasks and
othorgonal to our approach.

There have been many hardware proposals that allow communications
between the system software (OS) and the hardware to make better
resource scheduling/allocation decisions. For example,
many DRAM controller design proposals allow the OS to set priorities, on a
per-core basis, on memory request
scheduling~\cite{kim2010atlas,kim2010thread,subramanian2013mise,iyer2007qos}. 
More recently, Intel's new Xeon architecture
begins to expose shared resource allocation interfaces, currently
restricted to partitioning the LLC but the interface is generic which
can support controlling other shared resources such as DRAM, to the OS~\cite{intelswref}
Such hardware support can be especially useful for BWLOCK 
because the software based periodic bandwidth control mechanism can be
replaced by more efficient hardware mechanisms with lower overhead. 

\section{Conclusion} \label{sec:conclusion}

We have presented BWLOCK, a user-level API and kernel-level memory bandwidth control
mechanism, designed to protect performance of soft real-time
applications such as multimedia applications. It provides simple lock
like APIs to declare memory-performance critical sections in the
application code. When an application accesses a memory critical
section, BWLOCK automatically regulates the other cores' so that they
cannot cause excessive memory interference.

We applied BWLOCK in two real-world soft real-time
applications---Mplayer and WebRTC framework---to protect
their real-time performance in the presence of memory intensive non
real-time applications that share the same machine. In both cases, we
were able to achieve near perfect real-time performance, or to choose
not perfect---but still better than the vanilla Linux---real-time
performance for minimal throughput reductions of non-real-time
applications.

Our future work includes hardware assisted bandwidth control for
better control quality and compiler based automatic
identification of memory-performance critical sections in soft
real-time applications.

\section*{Acknowledgements} \label{acknowledge}
This research is supported in part by NSF CNS 1302563. Any opinions,
findings, and conclusions or recommendations expressed in this
material are those of the authors and do not necessarily reflect the
views of the NSF.

\bibliographystyle{plain}
\bibliography{heechul}
\end{document}